# Ultrasonic study of the gelation of gelatin: phase diagram, hysteresis and kinetics

**N. G. Parker[1] and M. J. W. Povey**
School of Food Science and Nutrition, University of Leeds, Leeds, LS2 9JT, UK

E-mail: n.p@physics.org

**Abstract.** We map the ultrasonic (8 MHz) speed and attenuation of edible-grade gelatin in water, exploring the key dependencies on temperature, concentration and time. The ultrasonic signatures of the sol-gel transition, confirmed by rheological measurements, and incomplete gel formation at low concentrations, enable a phase diagram of the system to be constructed. Sensitivity is also demonstrated to the kinetics of gel formation and melting, and associated hysteresis effects upon cyclic temperature sweeps. Furthermore, simple acoustic models of the sol and gel state enable estimation of the speed of sound and compressibility of gelatin. Our results demonstrate the potential of ultrasonic measurements to characterise the structure and visco-elasticity of gelatin hydrogels.

## 1. Introduction

A hydrogel is a colloidal phase of matter comprising of a solid network of macromolecules or particles interspersed with an aqueous phase which constitutes up to 99% of the total volume. Gels are further characterised by their solid-like mechanical abilities to maintain their shape in the absence of external stress. Gelatin is the quintessential gel, and whose special properties have been exploited as far back as ancient Egypt (David, 2007). Nowadays, gelatin hydrogels are extensively employed in food science for the purposes of thickening and textural engineering, adhesion, emulsification, and stabilization of foams and films (Nishinari, et al., 1993; Ward & Courts, 1977, Imeson, 1997). Such capabilities have also found major applications of gelatin in the cosmetics, pharmaceutical and biomedical industries (Peppas, 1986). In food applications, the functional quality of gelatin is largely reliant on its visco-elastic properties. The inherent variability of gelatin, arising from the variability of the originating biological tissues, places strong emphasis on characterisation techniques. Moreover, the current trend towards fish-based gelatins is hampered by their particularly large variability, prompting the need for advances in both the engineering and quality assurance of gel rheology (Karim & Bhat, 2008).

Gelatin is derived from collagen, the fibrous protein that forms a major component of mammalian flesh and connective tissues. Collagen exists in nature as a macromolecule of three polypeptide strands, approximately 300 nm long, contorted into a triple helix conformation, and with a molecular weight of around 100kDa (Djabourov, Leblond, & Papon, 1988a; Guenet, 1992). Under hydrolytic degradation (by simply boiling collagen-rich tissue or more sophisticated acidic and alkaline preparation), the collagen triple helix separates into the three polypeptide strands that compose gelatin. Gelatin has the ability to form a homogeneous gel for concentrations in the range of approximately 1-50% (below this range there are insufficient molecules to support an infinite three-dimensional gel network) (Djabourov, et al., 1988a; Guenet, 1992). The phase diagram of aqueous gelatin has been established through calorimetry and optical approaches (Bohidar & Jena, 1993; Djabourov, et al., 1988a). Above the gelling temperature, typically around 30°C, a gelatin-water mixture exists as a sol. For sufficiently high temperatures (greater than approximately 50°C) the dispersed particles are gelatin monomers. As the temperature is lowered the gelatin monomers aggregate via hydrogen bonding into oligomers which snake through the liquid in a random coil conformation. At lower temperature, the increasing role of intra-oligomer hydrogen-bonding induces a transition from disordered coils to ordered single helices. Finally, the sol-gel transition occurs, driven by a crossover from single to triple helices. The triple helices reorganize over time to form an equilibrium infinite gel network (Bohidar, et al., 1993) in a manner consistent with percolation theory (Djabourov, et al., 1988a). Finally, a giant interconnected macromolecular network is formed in which the triple helices form the junctions that bind the gel network and provide its strength and elasticity. Note that the formation of triple helices is a partial renaturation of the origin collagen state. Since the bonds that drive this transition are physical (hydrogen and van der Waals bonds), the gelation process is thermo-reversible. Indeed, it is the relatively low energy required to break these bonds that provides gelatin gels with their touted thermo-physical properties.

---

[1] To whom any correspondence should be addressed.

Gelation of gelatin is most commonly monitored through calorimetry (sensitive to heat-induced structural changes), optical rotation measurements (sensitive to the degree of helical conformation) and mechanical rheometry (sensitive to low-frequency visco-elastic properties) (Guenet, 1992). Ultrasound provides an alternative and complimentary approach to study fluid-based systems, buoyed by its sensitivity to a material's viscous and elastic properties down to the micro-scale (Povey, 1997). As such ultrasound has found applications in studying phase transitions, e.g., the first-order transition of oil crystallization in emulsions (Povey, 2000) and second-order glass-rubber transitions in polymers (Bordelius, Troitskaya, & Semenova, 1973; Parker, Mather, Morgan, & Povey, 2010; Smith & Wiggins, 1972). Further attractive features of ultrasound are its ability to probe non-invasively, with negligible sample degradation, to access many opaque materials, and to be extended to industrial scales.

Longitudinal ultrasound, in which the particle displacements are parallel to the direction of propagation, is most commonly used and will be considered here. By definition, ultrasound has frequency greater than 20 kHz. The speed of sound $v$ and attenuation $\alpha$ through the medium are the measureable parameters of interest. Assuming a Newtonian liquid with negligible thermo-acoustic losses, these parameters can be directly related to the complex longitudinal elastic modulus $M=M'+iM''$ via (Povey, 1997),

$$M' = \rho v^2, \qquad (1)$$

$$M'' = \frac{2\rho\alpha v^3}{\omega} \qquad (2)$$

where $\rho$ is the material density and $\omega$ is the angular frequency of the wave. The link to the conventional rheological parameters - the bulk $K$ and shear moduli $G$ - is given by $M=K+4G/3$. In typical hydrogels the bulk modulus is dominated by the water phase and of the order of GPa. In contrast, the shear modulus is dramatically increased by the gel network: while water has a shear modulus of order $10^{-5}$Pa, that of gelatin gels is of the order of kPa (Ziegler & Rizvi, 1989). Still, this remains much less than the bulk modulus, and we can expect that changes in the shear modulus will have negligible effect on the speed of longitudinal sound. At 8 MHz, the ultrasound wavelength $\lambda=2\pi v/\omega$ in gelatin will be approximately 0.25 mm. Given that the typical diameter of a gelatin filament is ~1 nm and the typical inter-filament distance is ~5 nm, such ultrasound will probe the bulk, collective properties of such structures.

Ultrasound has been employed to study gels and the sol-gel transition within, for example, biopolymers (Audebrand, Doublier, Durand, & Emery, 1995; Burke, Hammes, & Lewis, 1965; Emery, Chatellier, & Durand, 1986; Toubal, Nongaillard, Radziszewski, Boulenguer, & Langendorff, 2003; Wada, Sasabe, & Tomono, 1967), epoxies (Sidebottom, 1993) and blood plasma (Calor & Machado,

2006).  On the theoretical side, several models of sound propagation in gels exist (Bacri, Courdille, Dumas, & Rajaonarison, 1980; Johnson, 1982; Marqusee, et al., 1981).   Only a handful of experimental studies have specifically considered the ultrasonics of gelatin (M. Audebrand, M. Kolb, & M. A. Axelos, 2006a; Bacri, et al., 1980; Choi & Tanimoto, 2000; Emery, et al., 1986; Toubal, et al., 2003; Zhao & Vanderwal, 1997).   As such we shall also draw from ultrasonic studies of the closely related polysaccharides, as well as hypersonic studies of gelatin performed via Brillouin light scattering (Bot, et al., 1995; Zhao, et al., 1997).  A general trend is that the sound attenuation undergoes a marked change across the sol-gel transition, even for low concentrations of the order of 1% (M. Audebrand, M. Kolb, & M. A. V. Axelos, 2006b).  The sound attenuation has been shown to be strongly correlated with rheological and turbidity measurements, and potentially offer a more sensitive probe of gelation (Audebrand, et al., 1995; Audebrand, et al., 2006a).

The sensitivity of the sound attenuation to the gel state arises from structural and viscous relaxation effects in the gel (Eggers & Funck, 1976).  When these effects are stimulated at close to their resonant frequency (which fortuitously typically coincide with readily-achievable ultrasonic frequencies), a significant amount of sound energy becomes dissipated.   This possibility has motivated ultrasound spectroscopy of relaxation phenomena in gelatin and polysaccharides (Audebrand, et al., 1995; Audebrand, et al., 2006b; Bacri, et al., 1980; Burke, et al., 1965; Choi, et al., 2000) to, for example, probe hydration (Choi, et al., 2000) and the helix-coil transition (Burke, et al., 1965; Emery, et al., 1986).

Beyond this, little information of the ultrasonic behaviour of the gelatin-water system has been established.  Their remains no observation that the gel-sol transition affects the sound speed in gelatin, in contrast to the marked changes in polysaccharide gels (Toubal, et al., 2003). Meanwhile, there is a paucity of data on how the speed of sound and attenuation scale with gelatin concentration.  For example, Bot *et al*. (1995) found the speed of sound to increase as the square root of the gelatin concentration (in the hypersonic regime), while Bacri *et al.* (1980) found the sound attenuation to increase with concentration-squared.  The potential application of ultrasound to monitor gelatin processes and quality requires advanced knowledge of their ultrasonic signatures and available sensitivity. Motivated by this, we here map out the behaviour of the sound speed and attenuation of gelatin-water samples (made of a typical food-grade commercial gelatin) as a function of temperature, concentration and time. We consider temperature over the range 5-80$^\circ$C, from well-developed infinite gels to highly dispersed sols, and concentrations from 0.5% up to 30%, from the critical limit for gel homogeneity up to strong gels.  Furthermore, we explore the identification of the gel-sol transition, and the kinetics and reversibility across it.   Where appropriate, rheological measurements are performed to corroborate and aid in interpreting our results.

## 2. Experimental

*2.1. Gelatin and preparation of aqueous samples*
Edible-grade gelatin (acid type, 200 Bloom, ash content < 2%), derived from bovine tissue, was supplied by Geltech Co. (Busan, Korea).  After adding the required concentration of gelatin (by weight) to Millipore water (0.22 micron filter), each mixture is stirred at 60 $^\circ$C for 60 minutes to form a visibly homogeneous sol. Heating for 30 minutes above 40-50$^\circ$C has been shown to form a gelatin sol in its monomer state and erase its thermal memory (Guenet, 1992), and we expect to be well within this regime. Weight concentrations of 0.1, 0.5, 1, 2.5, 5, 7.5, 10, 15, 20, 25 and 30 % are employed, as well as pure water for comparison.  Higher concentrations of gelatin were not possible since the solution became too viscous to be employed in the instrumentation.  Each gel is studied for no longer than 18 hours following preparation during which bacterial growth will be negligible.

*2.2. Ultrasound measurements*
Ultrasound measurements were performed using a Resoscan instrument (TF Instruments, Heidelberg, Germany).  The device contains two sample cells (200 μl capacity) which act as ultrasound resonators with a path length of 7 mm and operating frequency range 7.3-8.4 MHz. By identifying the resonant frequency of the cell, the instrument determines the sound speed and attenuation in the sample (with a repeatability of 0.01 ms$^{-1}$ and a few percent, respectively, according to the manufacturer). A programmable Peltier thermostat allows the sample temperature to be varied in range 5-85 $^{o}$C with an absolute accuracy of ±0.05 $^{o}$C (manufacturer's specification). The instrument is set to gather ultrasonic data every 10 seconds. The presence of the two cells allows for differential measurements, although here we employ the second cell for consistency checking and averaging.

A verification measurement with Millipore water gave excellent agreement with the known speed of sound across the whole temperature range of the instrument. Following each gelatin measurement the sample cells are heated to 80 $^{o}$C and flushed repeatedly with water.

*2.3. Ultrasonic measurement protocols*

*2.3.1. Constant temperature*
The variation of the ultrasonic parameters with gelatin concentration is assessed at fixed temperatures of 5, 20 and 80 $^{o}$C. Hot gelatin solution (~60 $^{o}$C) is pipetted into the Resoscan cells, which are set to 80$^{o}$C to promote flow into the cells. The sample is ramped slowly (0.5 $^{o}$C min$^{-1}$) to the desired temperature and aged for 30 minutes, before taking ultrasonic measurements over a one hour period.

*2.3.2. Heating-cooling cycles*
To examine the variation of the ultrasonic behaviour with temperature we perform heating-cooling cycles across the gel-sol transition. After filling, the sample is maintained at 80 $^{o}$C for 30 minutes. The temperature is then ramped down to 5 $^{o}$C at a rate of -0.5 $^{o}$C min$^{-1}$ and maintained for a further 30 minutes. The temperature is then ramped back up to 80 $^{o}$C to complete the cycle. This cycle is repeated once more to verify repeatability.

*2.3.3. Quenching through the gel-sol transition*
To probe the gelatin kinetics, we quench the sample from above the gel transition (60 $^{o}$C) to below it. The Resoscan device is set to the required final temperature and hot solution (~60 $^{o}$C) is pipetted in. Ultrasonic measurements are then taken continuously for three hours. In the first set of measurements, the concentration is set to 5 % (w/w) and final temperatures of 5, 10 20 and 30 $^{o}$C are considered. Secondly, with the final temperature fixed to 20 $^{o}$C, different concentrations are considered: 0, 0.5, 1, 2.5 and 5 %. Note that the variation of gelatin density/specific volume with temperature is approximately the same as water (Davis & Oakes, 1922): over a quench from 60 $^{o}$C to 20 $^{o}$C the volume contracts by less than 2 %. Thus, we do not expect that the sample will noticeably retract from the face of the ultrasound transducers (which would cause anomalous results).

*2.4. Rheological measurements*
We probe the visco-elastic properties using a temperature-controlled Kinexus rheometer (Malvern, UK) in cone-plate arrangement (60 mm diameter cone with 2$^{o}$ angle and a 65 mm diameter plate). A vibratory rotational strain with amplitude $\varepsilon_0$ is applied to the sample, generating a vibratory stress response of amplitude $\sigma_0$. In the linear visco-elastic regime, the complex shear modulus is provided by their ratio,

$$G = G' + iG'' = \frac{\sigma_0}{\varepsilon_0}(\cos\delta + i\sin\delta) \qquad (3)$$

where $\delta$ is the phase lag between the stress and strain oscillations. We employ a fixed oscillation frequency of 1 Hz. Approximately 2 ml of molten gelatin is applied to the plate at elevated temperature (~60 $^{o}$C). The temperature is then changed to the required value and, after 30 minutes of aging, measurements are performed. We perform both strain sweeps and temperature sweeps. In the former case, the applied stress is varied between $10^{-4}$ and 1 Pa. In the latter case, the shear strain is fixed at 0.05 %.

## 3. Results and discussion

### 3.1. Concentration-dependence of acoustic parameters

Following the protocol of Section 2.3.1, we determine the average (over one hour) speed of sound and attenuation at 5, 20 and 80 °C as a function of gelatin concentration $C$ (w/w). First consider the speed of sound (Figure 1(a)). All three temperatures shows the same qualitative behaviour: the speed of sound increases with concentration in an approximately linear manner, with the speed greatest at the highest temperature. Over the concentration range considered, the speed of sound varies by up to 10 %. Next consider the attenuation (Figure 1(b)). The attenuation increases faster-than-linear with concentration, in agreement with (Bacri, et al., 1980). However, whereas Bacri *et al.* suggest a quadratic dependence on $C$, we do not observe any single power law scaling of the data. The attenuation markedly increases with decreasing temperature. Indeed, the considerably high attenuation at 5 and 20 °C can be attributed to the presence of the gel state, in accord with previous observations (Bacri, et al., 1980; Choi, et al., 2000).

At 80 °C, the system will be in the sol state. In contrast, the 5 and 20 °C measurements are below the gel-sol transition temperature. However, the formation of a gel state cannot be assumed at the lower end of the concentrations considered. It is established that below a critical concentration, typically around 1 %, there is insufficient gelatin to support an infinite gel network. This is the origin of the discontinuity in the ultrasonic speed and attenuation at $C \approx 1.75$ %. This region is most apparent in the inset of Figure 1(b). Here the speed of sound flattens off as one approaches $C=0$, indicating a vastly reduced sensitivity to the presence of gelatin. In support of this, there is no such discontinuity in the 80°C sol system, and we additionally observe that for concentrations up to 1.75 % an infinite gel does not visibly form at room temperature.

### 3.1.1. Modeling the sol phase

The 80 °C data is the most readily interpreted since the system consists of a dispersion of gelatin monomers. The speed of sound of a pure isotropic fluid is given by neglecting the shear modulus in

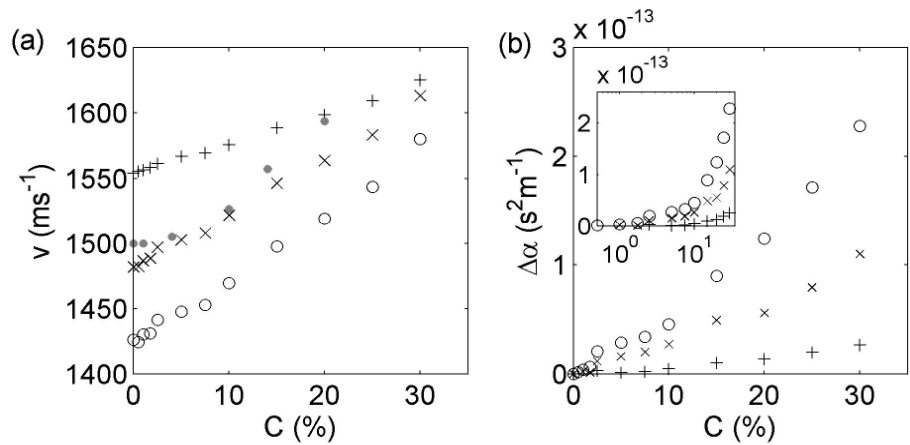

Figure 1: (a) Sound speed and (b) attenuation (relative to water) as a function of gelatin concentration (w/w) at 5 °C (circles), 20 °C (crosses) and 80 °C (pluses). The inset of (b) presents the same data on a logarithmic *x*-scale. Measurement errors in concentration are less than 1% of the absolute value. Each ultrasound data point is the average over one hour; during which the speed of sound varies by less than 0.1 ms$^{-1}$ and the attenuation by less than $2 \times 10^{-15}$ s$^2$m$^{-1}$. In (a) the hypersonic data of (Bot, Schram, & Wegdam, 1995) is shown for comparison (grey dots).

Eq. (1),

$$v = \sqrt{\frac{1}{\kappa\rho}} \qquad (4)$$

where $\kappa = K^{-1}$ is the bulk compressibility. To apply this to our two-component system we employ the Urick model (Povey, 1997). Here the system is treated as an effective homogeneous medium with a compressibility and density given by the volume average of the component systems,

$$\kappa = (1-\phi)\kappa_w + \phi\kappa_g$$
$$\rho = (1-\phi)\rho_w + \phi\rho_g \qquad (5)$$

where $\phi$ denotes the volume fraction of gelatin, $\kappa_w$ and $\kappa_g$ are the compressibilities of water and gelatin, and $\rho_w$ and $\rho_g$ are the densities of water and gelatin. The parameters for water at 80 °C are obtained from the literature to be $\rho_w$=971.82 kg m$^{-3}$ and $\kappa$=4.267×10$^{-10}$ m$^2$ N$^{-1}$ (Kaye & Laby, 1995; Lide, 2003). We determine the density of gelatin by measuring the density of gelatin-water mixtures with concentration (see Appendix A). In principle, the mixture density should be related to the partial molar volume of the gelatin and water to allow for 'excess volume' effects arising from the interaction between the solute and solvent (Atkins & De Paula, 2006). However, we find that the partial molar volume does not significantly change with concentration (as indicated by the linear dependence on concentration in Appendix A) and thus proceed under the assumption of additive densities, as in Equation (5). From the density measurements we determine $\rho_g$=1332.6 kg m$^{-3}$. The remaining unknown is the compressibility of gelatin, to be determined by fitting. The Urick model gives very good agreement with the experimental speed measurements, as illustrated in Figure 2(a). From this we determine the compressibility of gelatin to be 1.412×10$^{-10}$ m$^2$ N$^{-1}$, and accordingly the speed of sound of gelatin is $v_g = (\kappa_g \rho_g)^{-1/2}$ = 2305.2 ms$^{-1}$.

*3.1.2. Modeling the gel phase*

We will now focus on the data corresponding to a homogeneous gel data ($T$=5 and 20 °C, and $C$>1.75 %). The only previous explicit measurement of speed of sound versus concentration was performed by Bot *et al.* in the hypersonic regime, and this data is presented by grey dots in Figure 2(a). It is interesting to note that their data also indicates a discontinuous change at low concentration. Above this, their data also scales approximately linearly with concentration but has a considerably larger gradient. Differences are to be anticipated: for hypersonic studies the sound has sub-nanometer wavelengths and will directly probe the nanoscale gelatin filaments and their mesh-structure, whereas ultrasonic waves will probe the bulk, averaged structure.

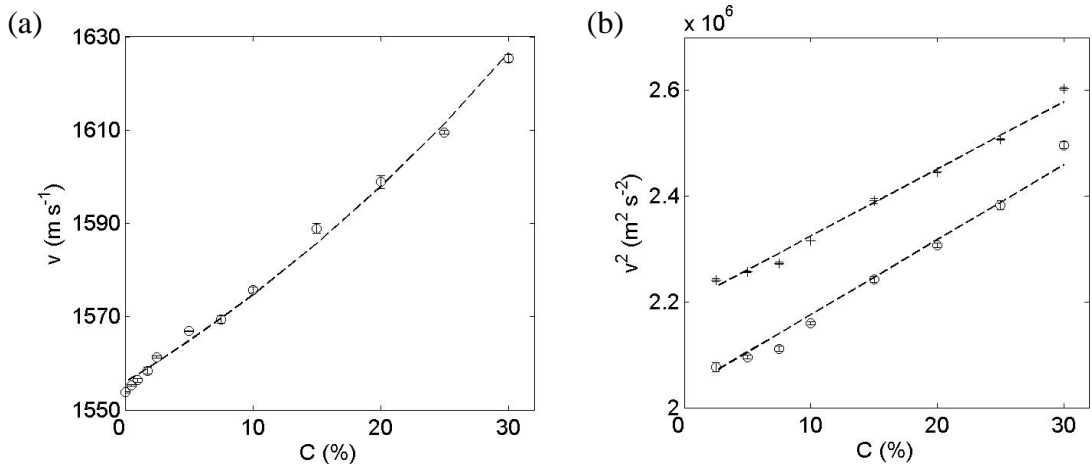

Figure 2: (a) Speed of sound at 80°C as a function of concentration. The dashed line shows the Urick model with gelatin compressibility $\kappa$=1.412×10$^{-10}$ m$^2$ N$^{-1}$ and density $\rho$=1332.6 kg m$^{-3}$. (b) Speed of sound at 5 (circles) and 20 °C (squares), along with their respective fits to the model of Marqusee and Deutch. The fitting parameters are $v_n$=1862.1 ms$^{-1}$ and $\lambda$~10$^{-7}$ at 5 °C, and $v_n$=1857.5 ms$^{-1}$ and $\lambda$~10$^{-7}$ at 20 °C.

Marquesee and Deutch (1981) have developed a hydrodynamic model for sonic propagation through gels that incorporates the effects of frictional and elastic coupling between the fluid and network. In the case of strong friction in both the fluid and network, the model reduces to an average coupled elastic medium. Re-expressing equation (4.19) of Marqusee, et al. (1981) in terms of weight concentration $C$, we obtain the following equation for the gel speed of sound,

$$v^2 = v_w^2(1-C) + v_n^2 C + 2 v_w v_n \sqrt{\lambda C(1-C)}. \tag{6}$$

Here $v_n$ is the speed of sound in the gel network, which may deviate from the gelatin speed of sound in the sol state determined earlier, and $\lambda$ is an elastic coupling parameter which scales from no ($\lambda=0$) to maximal coupling ($\lambda=1$). Both $v_n$ and $\lambda$ are used as fitting parameters, while $v_w$ is taken from the literature. We obtain very good fits to the speed of sound data, as shown in Figure 2(b). For 5 °C and 20°C, the gelatin network speed of sound is 1862.1 and 1857.5 ms$^{-1}$ respectively. In both cases, $\lambda$ is of the order of 10$^{-7}$, suggesting that there is practically no elastic coupling between the fluid and network. Consistent with this, Bacri *et al.* successfully employed a hydrodynamic model without an elastic coupling term and found good agreement with ultrasonic data. In contrast, when performed in the hypersonic regime, elastic coupling is found to be present; Bot *et al.* determine a value of $\lambda=0.333$.

*3.2. Variation with temperature*

Figure 3(a) plots the variation of the speed of sound during a heating phase from 5 to 80°C for various concentrations of gelatin. The speed of sound follows similarly shaped curves in all cases, first increasing with temperature, reaching a peak in the vicinity 50-75°C, and then decreasing with temperature. This shape arises from the water component of the sample. For most fluids the speed of sound decreases monotonically with temperature since thermal energy leads to a monotonically increasing compressibility and decreasing density (see Equation (4)). However, water is anomalous: as temperature is decreased, the molecules adopt form 'open' quasi-crystalline aggregates (Kell, 1975), causing the compressibility to increase and induce a maximum in the speed of sound at approximately 74°C (for pure water). The speed of sound curves shift upwards with increasing concentration and the peak shifts towards lower temperatures. There is a significant variation with concentration across the whole temperature range. However, no obvious signature of gelatin is evident in these speed curves.

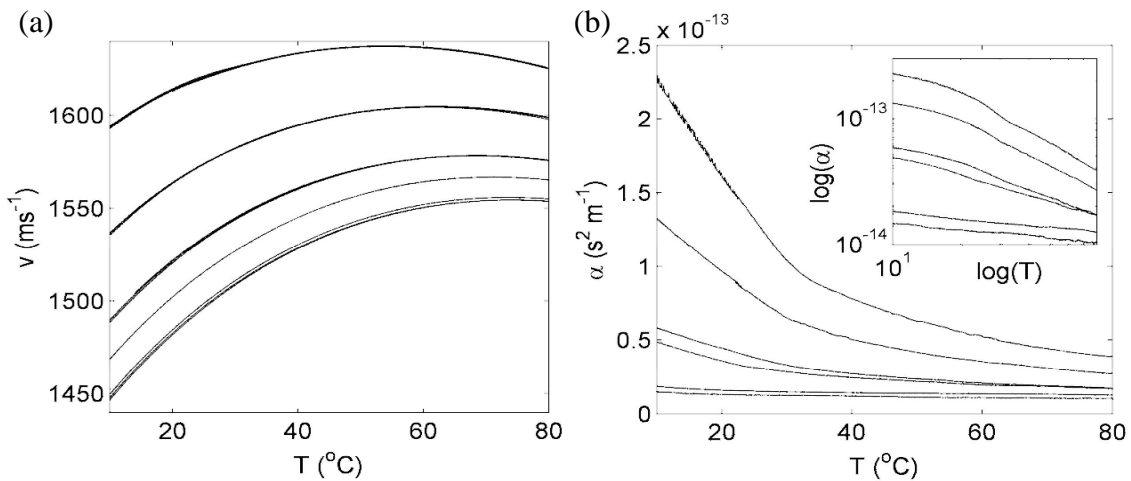

Figure 3: (a) Speed of sound and (b) attenuation as a function of temperature (during a heating sweep). In (a) the curves correspond to concentrations of, from top to bottom, $C$=0, 0.5, 5, 10, 20 and 30%. In (b) the curves correspond to, from bottom to top, C=0.5, 0, 5, 10, 20 and 30%. NB The anomalous ordering of the 0 and 0.5% cases in (b) is indicative of the very weak attenuation observed in both cases.

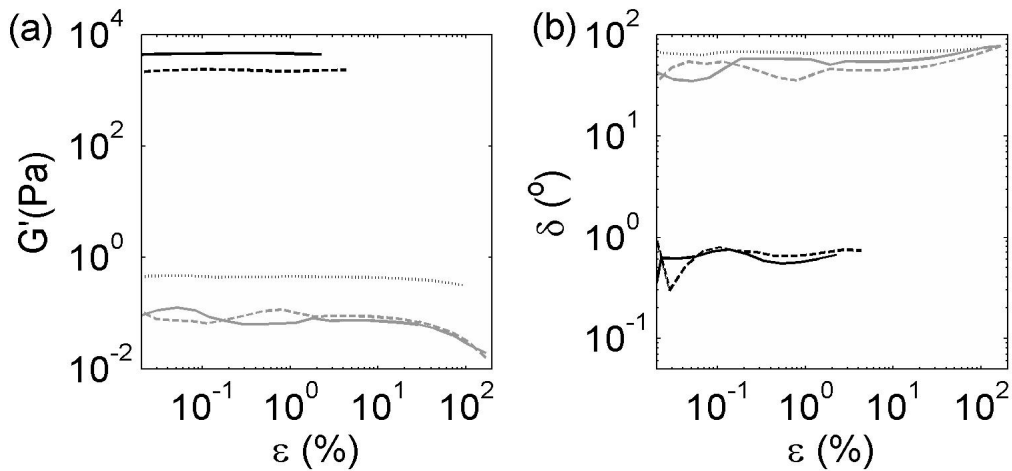

Figure 4: (a) Storage modulus and (b) phase angle as a function of shear strain of a 10% (w/w) gelatin system. The measurements are taken at various temperatures (°C): 10 (solid black line), 20 (dashed black line), 30 (dotted black line), 40 (solid grey line) and 60 (dashed grey line). NB The data is taken over a stress range $10^{-4}$-1 Pa, hence the strain range varies between temperatures.

The corresponding plots of the sound attenuation are plotted in Figure 3(b). The general trend is that the attenuation decreases with increasing temperature. In general, for pure fluids, the ultrasound attenuation decreases smoothly with temperature due to decreasing fluid viscosity (Holmes, Parker, & Povey, 2010). We see such smooth behaviour for the case of pure water and 0.5% concentration (most readily seen in the log-log plot (inset of Figure 3(b))). However, for $C$=5% and above there clearly exist kinks in the attenuation in the vicinity of 30°C which we attribute to the gel-sol transition. As the gelatin concentration increases the kink becomes increasingly apparent and the attenuation becomes larger. The increased sound attenuation in the gel phase appears to be a generic occurrence, and has been observed in gelatin (Bacri, et al., 1980) and polysaccharides (Audebrand, et al., 1995; Audebrand, et al., 2006a). In the gel state the attenuation has a steeper gradient than in the sol state.

We have additionally performed rheological measurements. Note that the generic rheological behaviour of gelatin-water systems has been previously studied (Djabourov et al., 1988b). Figure 4 presents strain sweeps of the storage modulus G' and phase factor δ at various temperatures for a 10% (w/w) system. The curves are approximately horizontal up to strains ε~10%, indicating the linear visco-elastic behaviour in this range. Deep in the gel phase (solid and dashed black lines), the storage modulus is large and the stress-strain oscillations are approximately in phase (the phase angle is close to zero), both of which are rheological characteristics of solid-like behaviour. In the sol phase (dashed and solid grey lines), the storage modulus drops dramatically and the strain and stress become out of phase (phase angle ~ 90°), consistent with the behaviour of a dominantly viscous fluid. Note that the rheological state of the system is thought to be largely determined by the number of triple helices in the system (Djabourov et al., 1988b).

*3.3. Hysteresis and thermoreversibility*
To investigate the thermoreversibility of the gel-sol transition we conduct temperature cycles as outlined in Section 2.3.2. Figure 5 presents the corresponding data for a 30% (w/w) sample. Although hard to visualise, there exists a small hysteresis loop in the speed of sound data in the vicinity of 25 °C (Figure 5(a)). This is more evident in the difference between the heating and cooling curves, $\Delta v$ (inset). The hysteresis loop extends from approximately 22 °C to 30 °C, with an amplitude of around 1 ms$^{-1}$.

Hysteresis effects are much more apparent in the attenuation (Figure 5(b)). Well above the gel-sol transition the heating and cooling behaviour is identical. However, in the region of the sol-gel transition the cooling and heating data clearly separate, and is this further visualised through the temperature-derivative of the attenuation data (inset). Upon heating the acoustic data changes slope at a higher temperature than the cooling phase. The curves are separated by approximately 8°C but otherwise share the same main features, i.e. to a good approximation the heating curve is shifted by +8°C relative to the cooling curve.

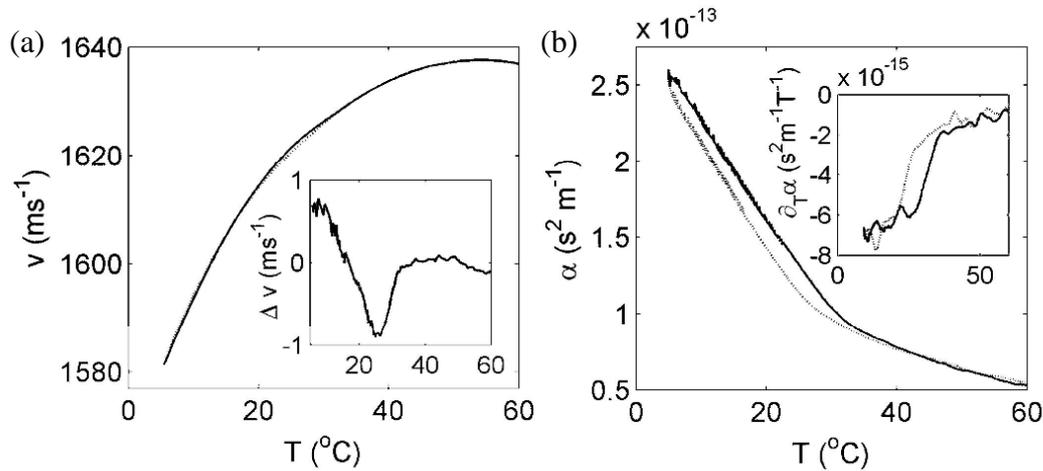

Figure 5: Ultrasonic behaviour during a temperature cycle of a 30% (w/w) gelatin solution. The black and dotted lines represent the heating and cooling cycles respectively. (a) Speed of sound, with the difference between heating and cooling curves (inset). (b) Attenuation, with its temperature derivative shown in the inset.

We note that these hysteresis effects in the speed of sound and attenuation are reproducible: for multiple temperature cycles the acoustic properties trace the same trend shown in Figure 5. The same qualitative behaviour is observed for all of the gel samples considered, but with the amplitude of the hysteresis effects rapidly diminishing as the gelatin concentration is reduced. We attribute this hysteresis to the different kinetics during the melting and gelation of gelatin, for example, as observed in optical rotation measurements (Djabourov, et al., 1988a) and rheological properties (see Figure 6). In the former case, the amount of optical rotation gradually increased upon cooling through the gel-sol transition. Since optical rotation is caused by helical structures, this established the kinetic process of the coil-helix transition that drives gelation: over a timescale of typically tens of minutes the network junctions assemble and disassemble until a suitable low energy state is achieved that extends across the whole system. It is only once this infinite gel network has formed that the ultrasonic properties significantly change form and enter their gel regime. In our measuremements, the temperature ramping timescale is faster than the kinetic timescale for network formation, preventing the system from following its thermal equilibrium states. In contrast the melting of the gel network occurs sharply once the thermal energy is sufficient to break the junctions, and will be largely independent of ramping time. Here, a distinct hysteresis is formed upon heating and cooling. It is intriguing to note that our temperature-attenuation curves mirror their temperature-optical rotation curves, suggesting a strong correlation between the sound attenuation and the proportion of helix formation.

*3.4. Determining the sol-gel transition: ultrasonic versus rheological approaches*
There is a marked change (a "kink") in the ultrasonic attenuation as the sol-gel transition is traversed, enabling the transition point to be determined. Since the attenuation is approximately linear above and below this kink, we will proceed by linear fitting in these regimes. We then define the transition point to be the intersection of these linear fits. This is analogous to the acoustic method used to estimate the glass transition temperature of polymers (Bordelius, et al., 1973; Parker, et al., 2010; Smith, et al., 1972). In Figure 6(a) we plot the attenuation in the vicinity of the sol-gel transitions for a 10% (w/w) sample. Linear fits clearly provide a good model of the behaviour adjacent to the transition. In the vicinity of the transition, the data and linear fits deviate over a region approximately 3°C wide, which is indicative of the width of the gel-sol transition itself. From the linear intersects, the transition temperatures are found to be 24.4 °C for gelation and 30.5 °C for melting.

A conventional approach to determine the sol-gel transition is through rheology measurements. In the gel phase, the storage modulus dominates while in the sol phase the loss modulus dominates. Thus, as the sol-gel transition is traversed, the loss and storage moduli must cross. This crossing point ($G'=G''$ or $\delta=45°$) is taken to define the transition point (Guenet, 1992; Djabourov et al., 1988b). We have additionally performed this approach. for the 10% (w/w) sample. In Figure 4 we saw that the linear visco-elastic regime existed up to strains of ~10%. Here we employ a strain of 0.05%, which is well within this linear regime. The storage and loss moduli are plotted in Figure 6(b), and clearly demonstrate the expected behaviour. The corresponding transition temperatures are 24.8°C and 30.2°C for gelation and melting, respectively, and in very good agreement with the ultrasonic measurements.

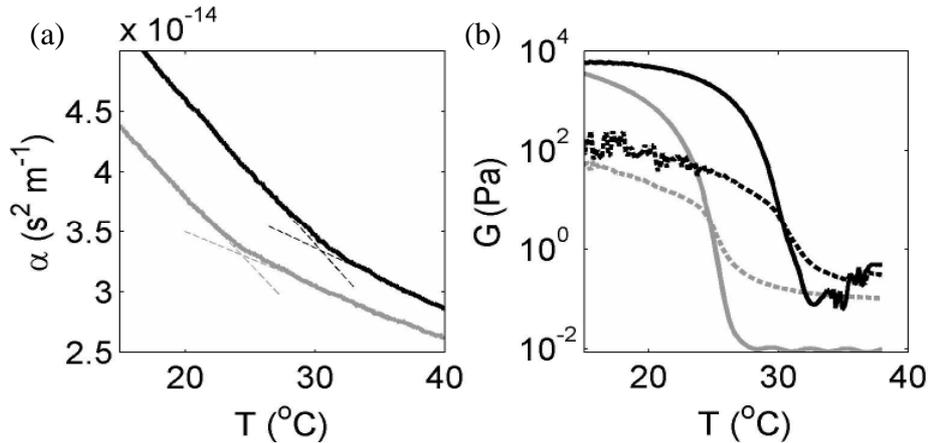

Figure 6: Ultrasonic and rheological behaviour across the sol-gel transition of a 10% (weight) mixture under heating (black lines) and cooling (grey lines). (a) Sound attenuation, with linear fits above and below the transition regions (dashed lines). (b) Storage modulus $G'$ (solid lines) and loss modulus $G''$ (dashed lines).

To assess repeatability of the ultrasonic approach, we determined the sol-gel transition in 8 separately made samples (all 10% (w/w)). We find that average gelation (melting) temperature to be 25.2 (30.9) °C with a standard deviation of 0.8 (0.8) °C.

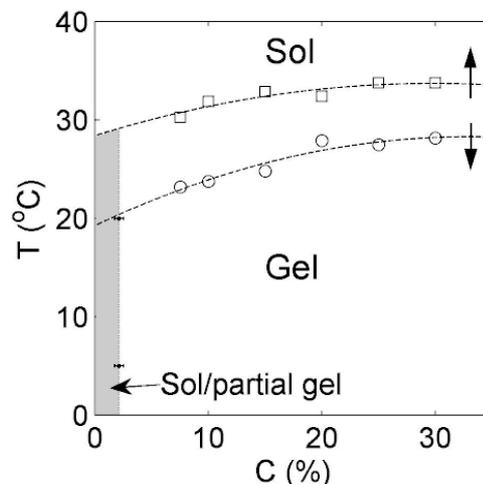

Figure 7: Concentration-temperature phase diagram of our aqueous gelatin sample showing the transition temperatures for melting (upper) and gelation (lower). The gelation line depends on cooling rate, which was 0.5 K min$^{-1}$ here.

## 3.5. Temperature-concentration phase diagram

We have determined the melting and gelation temperatures for different gelatine concentrations. In practice, for $C<7.5\%$ the kink does not appear clearly enough for the transition temperature to be determined with confidence. However, the results for concentrations of 7.5% and over are presented in Figure 7 where we form a temperature-concentration phase diagram for the state of the system. We additionally highlight the low concentration region in which the gel does not (at least within the timescale of the experiment) form an infinite network. This corresponds to where the acoustic data in Figure 1 undergoes a discontinuity as a function of concentration. The phase diagram is similar to that presented in (Bohidar, et al., 1993) from calorimetric analysis. However, while Bohidar et al. observe transitions within the sol phase (monomer-oligomer and coil-helix transition) we do not observe any ultrasonic evidence for these more subtle transitions.

## 3.6. Dynamics of the gelation process under quenching

To probe the kinetics of gelation we monitor the system following a sudden quench from well above the gel-sol transition (60°C) to lower temperature, as described in Section 2.3.3. Given the small sample volume (~200µl) and cooling power of the thermostat (90W, according to the manufacturer), we estimate the time to reach thermal equilibrium to be of the order of seconds. Even allowing for thermal losses, we can certainly expect thermal equilibrium within a few minutes.

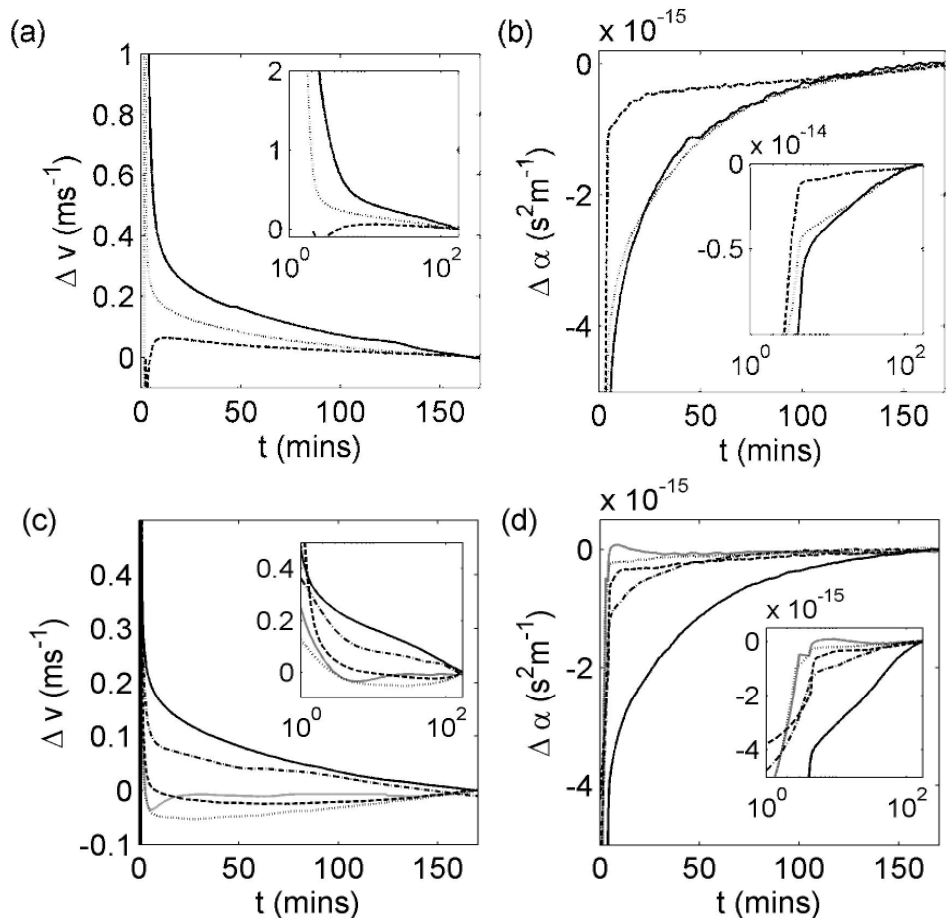

Figure 8: Change (relative to their final values) of (a) sound speed and (b) attenuation, during quenching of a 5% (w/w) gelatin solution from 60°C to final temperatures of 5 (dotted line), 10 (dashed line), 20 (dot-dashed line) and 30°C (solid line). (c) and (d) show similar plots but where the final temperature is fixed to 20°C and varying gelatin weight concentrations are employed: 0% (solid grey line), 1% (dotted line), 1.75% (dashed line), 2.5% (dot-dashed line) and 5% (solid black line). Each plot includes its corresponding log-linear plot as an inset.

Figure 8(a) and (b) presents the ensuing change in speed of sound and attenuation over a timescale of three hours for different final temperature (at fixed sample concentration of 5%). At very early times the speed of sound rapidly (~ minutes) decreases as the sample re-thermalizes to the lower temperature. Similarly, the attenuation rapidly increases during this time. For T≤20°C, the sound speed and attenuation then undergo a very fast decay over 10-20 minutes followed by a slow logarithmic decay. NB Because we plot the difference in sound speed and attenuation from their final values, the data sets tends towards zero. The observed scaling is similar to that observed in the degree of helical structures (via optical rotation measurements) following quenching (Djabourov, et al., 1988a). Based on this analysis we can attribute the short fast decay to the nucleation and growth of gelatin helical fibres and the slow logarithmic decay to reorganization (annealing) of the gel network. Even after three hours the ultrasonic properties of the samples still undergo logarithmic change, indicating the considerable timescale for annealing. For the 30°C case, the system appears to enter the slow logarithmic phase with no or minimal fast phase. This suggests the incomplete formation of gelatin fibres followed by slow reorganization processes, i.e., localised gelation of the sample.

Figure 8(c) and (d) present the ultrasonic properties following a quench for different gelatin concentrations (and a fixed final temperature of 20°C). We see marked ultrasonic differences between low/zero and high gelatin concentrations. For C=0, 0.5 and 1% there is essentially no change (including no obvious logarithmic scaling) in the speed of sound and attenuation following sample thermalization. This suggests that there are negligible reorganization processes occurring for C=0.5 and 1% and thus that these samples remain in the sol state (rather than undergoing localised gelation). In contrast, for C=2.5% and 5%, following thermalization there is the two-regime decay towards a fully formed gel state discussed above.

## 4. Conclusions

We have studied the ultrasonic characteristics of gelatin-water systems. We see that both the speed of sound and attenuation vary with concentration. This variation is consistent with simple acoustic models that enable the speed of sound and compressibility of gelatin to be determined. The sound speed and attenuation reflect the sol-gel transition and the transition to incomplete gelation at low concentrations, and enable the phase diagram for the system to be readily built up. The notable sensitivity of the ultrasound attenuation to the state of the system highlights the major role played by molecular relaxations. However, we observe no signs of the weaker monomer-oligomer and single-triple helix transitions that occur in the sol state.

We also observe hysteresis effects in the ultrasonic data arising from the kinetic processes during gelation, and specifically for gelation we observe two stage dynamics which we attribute to fast gelatin linkage followed by slow annealing.

Our results demonstrate that ultrasound provides a versatile approach to studying gelation, offering rapid data acquisition, insight into visco-elastic behaviour and the ability to probe small sample quantities. This may offer routes for non-invasive and non-destructive monitoring of industrial gelation processes. In particular, the ability of ultrasound to probe many opaque systems may hold particular relevance for studying more complex polymer systems which can additionally enter highly turbid chain-folded crystalline states (Guenet, 1992). Furthermore, the *fluctuations* in the transmitted sound waves may offer a route to probe the 'ghost' transient gel structures that precede full gelation and we hope to pursue this possibility in a future study

**Acknowledgements**
We thank Dr Mel Holmes for discussions and Professor Eric Dickinson for helpful comments in preparing this manuscript. We further acknowledge funding from the BBSRC (Ref: BBF0049231).

## Appendix A: Density of the gelatin-water dispersion

We perform density measurements of gelatin-water mixtures using a commercial density meter (Anton-paar DMA 4500 M). Measurements were taken at 80°C for which the system is a dispersion of gelatine monomers in water. The instrument measures density to an accuracy (repeatability) of $5\times10^{-5}$ ($1\times10^{-5}$) g cm$^{-3}$ and temperature to 0.03 (0.01) °C. At each concentration, five separate density measurements are taken. This was performed for concentrations up to 15% (beyond this the mixture was too viscous to be employ in the density meter). The experimental data is presented in Figure A.1 (points with error bars).

Assuming that the gelatin is homogeneously dispersed in the water, we model the data based on the additivity of densities (according to their volumes) encapsulated in Equation (5). It is trivial to show that the volume fraction $\phi$ of gelatin is related to the weight concentration via,

$$\phi = \frac{C\rho_w}{C(\rho_w - \rho_g) + \rho_g} \quad (A.1)$$

Taking $\rho_w$=971.8 kg m$^{-3}$ (Lide, 2003), we obtain an excellent fit to the experimental data that predicts the gelatin density to be $\rho_g$=1332.6 kg m$^{-3}$.

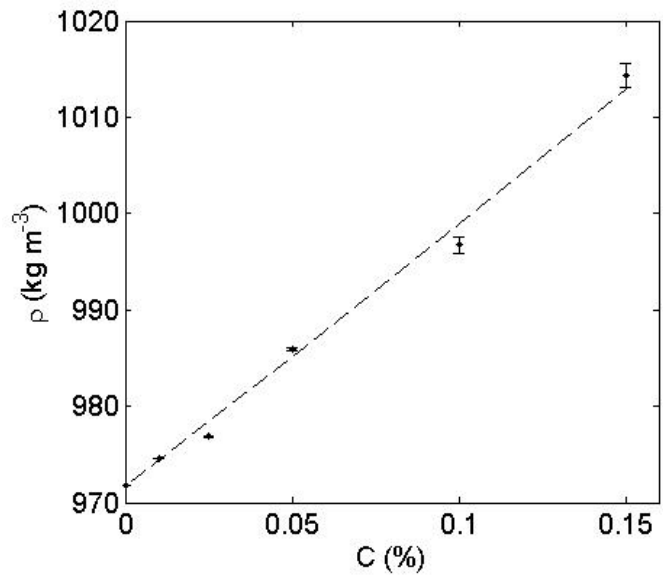

Figure A.1: Density of an 80°C gelatin-water mixture as a function of concentration (w/w). The dashed line is the least-squares fit of Equations (5) and (7) with $\rho_g$=1332.6 kg m$^{-3}$.